\documentclass[a4paper,12pt]{article}
\usepackage{amsmath}
\usepackage{amsfonts}
\usepackage{authblk}
\usepackage{graphicx,epsfig}

\newcommand{\BbbR}{\mathbb{R}}

\usepackage{color}

\numberwithin{equation}{section}

\title{Quantum memory for Rindler supertranslations}

\author{Sanved Kolekar$^{1,2,3}$\footnote{sanved.kolekar@cbs.ac.in} \,}
\author{Jorma Louko$^{1}$\footnote{jorma.louko@nottingham.ac.uk}}

\affil{$^{1}$ School of Mathematical Sciences, 
University of Nottingham,\\ 
Nottingham NG7 2RD, 
UK}

\affil{$^{2}$ CPT, Aix Marseille Universit\'e, Universit\'e de Toulon, CNRS, \\ UMR 7332,
13288 Marseille, France}

\affil{$^{3}$ UM-DAE Centre for Excellence in Basic Sciences,\\  
Mumbai 400098, India}

\date{Revised March 2018}
\begin{document}
\maketitle
\begin{abstract}
The Rindler horizon in Minkowski spacetime can be implanted 
with supertranslation hair by a matter shock wave without 
planar symmetry, and the hair is observable as a supertranslation 
memory on the Rindler 
family of uniformly linearly accelerated observers. We show 
that this classical memory is accompanied by a 
supertranslation quantum memory that modulates the entanglement 
between the opposing Rindler wedges in quantum field theory.  
A~corresponding phenomenon across a black hole horizon may 
play a role in Hawking, Perry and Strominger's proposal 
for supertranslations to provide a solution to the black 
hole information paradox. 
\end{abstract}

\section{Introduction}

In the long-standing pursuit to predict the fate of an evaporating black hole, 
a recent development is the suggestion by Hawking, Perry and Strominger 
\cite{HPS1,HPS2,Strominger:2017aeh} 
that significant quantum correlations may be encoded in 
``soft'' degrees of freedom, 
associated with vanishing energy in a particle description 
and a diffeomorphism in a geometric description, but nevertheless carrying 
nontrivial dynamics due to the global boundary conditions. 
Such degrees of freedom exist already in Minkowski spacetime 
electrodynamics~\cite{Kapec:2017tkm}, and in the gravitational case 
these degrees of freedom 
are associated with supertranslations in the Bondi-Metzner-Sachs (BMS) 
group of asymptotic isometries at the infinity
\cite{hotta1,hotta2,Strom-bms1,Winicour-memory,Ashtekar-memory,Strom-zhiboedov}. 
Contributions to the ongoing debate 
include~\cite{MP,Bousso:2017rsx}. 

In this paper we analyse the correspondence between quantum correlations and classical supertranslations 
in the simplified setting where a stationary black hole horizon is replaced by the Rindler horizon, the Killing horizon of a boost Killing vector in Minkowski spacetime. This simplification has a long pedigree~\cite{Fulling:1972md,Davies:1974th,unruh}, 
avoiding complications due to spacetime curvature but maintaining a bifurcate Killing horizon as a central piece of input in the quantum field theory \cite{Birrell:1982ix,Wald:1995yp}. We shall analyse how the quantum correlations across the Rindler horizon change when the horizon is implanted with classical supertranslation hair. 

Recall that the Schwarzschild black hole can be implanted with supertranslation hair by letting a spherically asymmetric shock wave fall into the hole~\cite{HPS2}. This classical hair is observable in the gravitational memory that affects the separation of geodesic observers at the asymptotic infinity~\cite{Zeld}, and in quantum field theory it is expected to be accompanied by correlations in the outgoing Hawking quanta. 

For a Rindler horizon, the notion of supertranslation hair has been characterised in \cite{Donn1,Donn2,Eling,Cai} (for a related discussion see~\cite{Hotta}). It was shown in \cite{san} that the Rindler horizon can be implanted with supertranslation hair by letting a shock wave without planar symmetry fall across the horizon, and this hair is classically observable in a memory on the Rindler family of uniformly linearly accelerated observers. We shall show that the classical Rindler supertranslation memory is accompanied by a Rindler supertranslation quantum memory, and we analyse how this memory modulates the entanglement between the opposing Rindler wedges. 

We work with a massless scalar field in $3+1$ spacetime dimensions. The core results are given in terms of a Bogoliubov transformation between a pre-supertranslation region and a post-supertranslation region, demonstrating that both the alpha-coefficients and beta-coefficients are nontrivial, so that the supertranslation induces both particle creation and mode mixing. The entanglement is analysed within a truncation to finitely many field modes, and using negativity as the entanglement monotone. We identify subsystems in which entanglement is degraded and subsystems in which the entanglement is generated, and this identification appears reasonably robust against the input used in the truncation. 

We anticipate that a similar analysis can be carried out for 
supertranslations implanted on a Schwarzschild black hole as in~\cite{HPS2} , 
and that the results will help to clarify the role of black hole 
supertranslations in the solution to the black hole information paradox. 

We begin in Section \ref{hairsection} with a recap 
of the classical Rindler supertranslation memory~\cite{san}. 
The quantum memory is found in Section~\ref{BG}, and the entanglement consequences
are analysed in Section~\ref{entanglement}. Section 
\ref{discsection} gives a summary and brief concluding remarks. 
Appendix 
\ref{app:K-Bessel-integral}
gives the derivation of an integral identity 
used in the main text, 
and Appendix \ref{app:negativity} recalls key features of 
negativity as an entanglement monotone. 

The Minkowski metric is taken to have the mostly plus sign, and
Roman indices run over all spacetime indices. 
Complex conjugate is denoted by an asterisk and Hermitian conjugate by a dagger.

\section{Recap: classical memory for Rindler supertranslations}\label{hairsection}

In this section we recall relevant properties of the Rindler 
spacetime and the classical 
Rindler supertranslation memory found in~\cite{san}, 
establishing the notation that will be used in the quantum field 
theory analysis in Section~\ref{BG}. 

\begin{figure}[t]
\includegraphics[width=\linewidth]{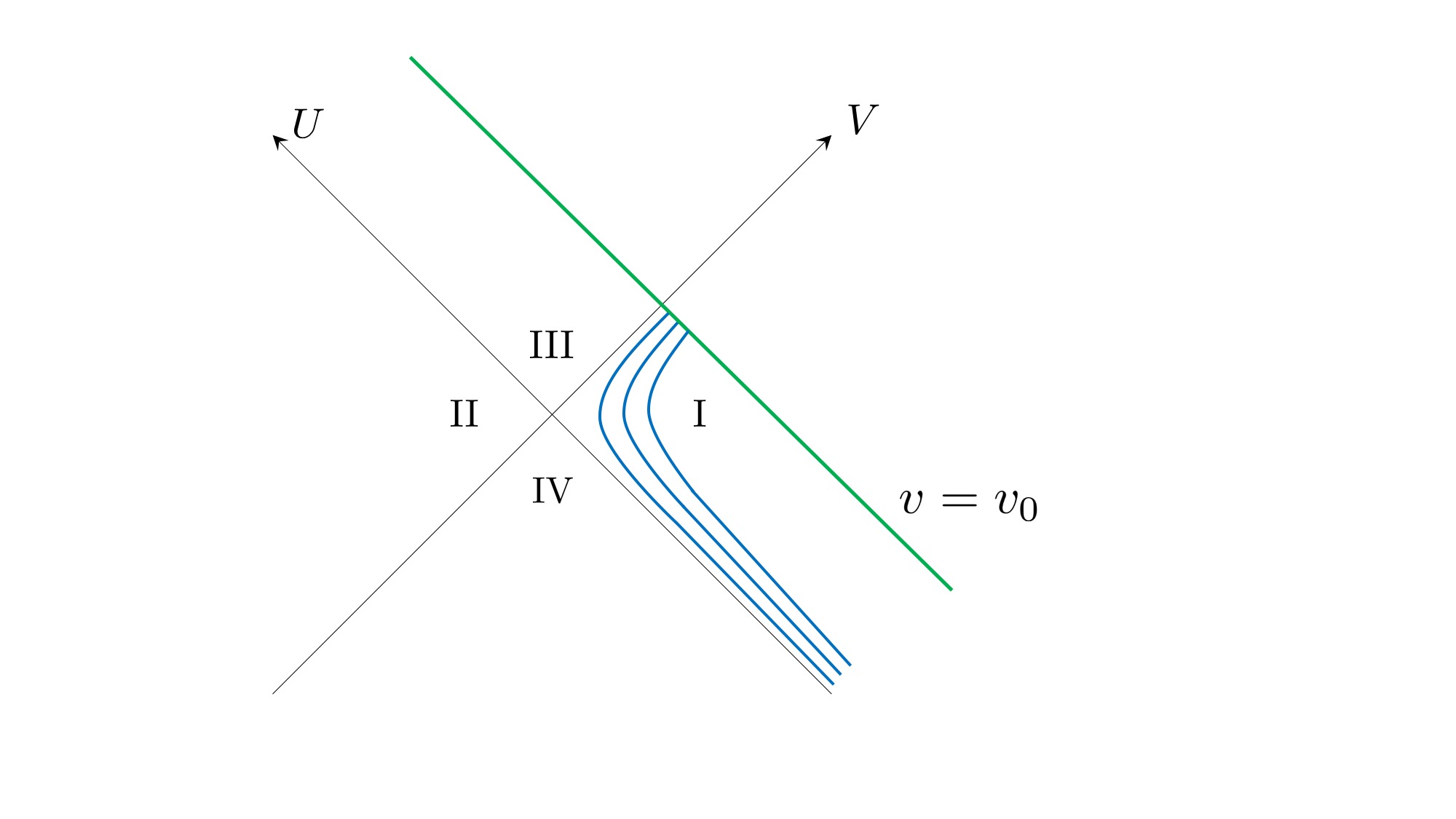}
\caption{A null shell in the Rindler spacetime, 
with the transverse dimensions suppressed. 
In the past of the shell the metric is given by~\eqref{eq:Minknullmetric}, 
and the shell is at $V = V_0>0$. The four Rindler quadrants are 
labelled in the figure by I, II, III, and~IV\null, and selected orbits 
of the boost Killing vector $\bar{\xi}$ 
\eqref{eq:boost-killing} are shown in quadrant~I\null. 
The coordinates in \eqref{eq:rindlershock1} cover both the future of the shell 
and the region $0<V<V_0$ of the past of the shell, 
and in these coordinates the shell is at $v=v_0$. 
\label{fig:diagram1}}
\end{figure}

The spacetime consists of two subsets of four-dimensional Minkowski spacetime 
joined together along a null shell as shown in Figure~\ref{fig:diagram1}. 
In the past of the shell we write the metric in 
the Minkowski null coordinates $(U,V,x,y)$ as 
\begin{align}
ds^2 = - dU\,dV + \delta_{AB}  dx^A dx^B
\ , 
\label{eq:Minknullmetric}
\end{align}
where the uppercase Latin indices take values in $\{x,y\}$ 
and the shell is at $V = V_0>0$. In the region $0<V<V_0$, 
we introduce advanced Bondi-type coordinates 
$(v,r,x,y)$ in which the metric reads 
\begin{equation}
ds^2 = - 2\kappa r dv^2 + 2 dvdr + \delta_{AB}  dx^A dx^B 
\ , 
\label{metricRindler}
\end{equation}
where $\kappa$ is a positive constant of dimension inverse length, 
$-\infty < r < \infty$, 
$-\infty < v < v_0$, and the shell is at $v=v_0$. 
In terms of the four Rindler quadrants shown in Figure~\ref{fig:diagram1}, 
$r>0$ is in region~I, $r<0$ is in region~III, 
and $r=0$ is on the Rindler horizon, $U=0$. 
Curves of constant $r$, $x$ and $y$ are orbits of the boost Killing vector 
\begin{align}
\bar{\xi} = -U\partial_U + V \partial_V  = \kappa^{-1} \partial_v
\ . 
\label{eq:boost-killing}
\end{align}
Selected orbits of $\bar{\xi}$ in region I are shown in the figure. 

In the future of the shell, $v>v_0$, 
we take the metric to be related to 
\eqref{metricRindler} by the diffeomorphism that is generated 
by the vector field 
\begin{equation}
\Xi^a = \kappa^{-1} \! \left[f(x,y), 0, -r \partial^A f(x,y) \right]
\ , 
\label{eq:Rindler-supertransfield}
\end{equation} 
where $f$ is an arbitrary function of the transverse coordinates. 
Working to linear order in~$f$, the metric for $-\infty < v < \infty$ thus reads 
\begin{align}
ds^2 &= - 2\kappa r dv^2 + 2 dvdr + 4r h(v-v_0) \partial_A f dvdx^A 
\nonumber 
\\[1ex]
& \hspace{3ex}
+ \left( \delta_{AB} + 2 \kappa^{-1} r h(v-v_0) \partial_A \partial_B f \right) dx^A dx^B 
\ ,
\label{eq:rindlershock1}
\end{align}  
where $h(v-v_0)$ is the 
Heaviside step function. 

While $\partial_v$ is not a Killing vector of \eqref{eq:rindlershock1}
at $v=v_0$, 
it is a Killing vector individually both for $v < v_0$ and for $v>v_0$, 
and in each region it generates a pure boost: 
for $v<v_0$ this holds by~\eqref{eq:boost-killing}, 
and for $v>v_0$ this holds because 
$\partial_v$ commutes with $\Xi$~\eqref{eq:Rindler-supertransfield}. 
As $g_{rr} = 0$ and $g_{vr} =2$, 
the coordinates in \eqref{eq:rindlershock1} may be regarded 
as a Rindler counterpart of Bondi-type coordinates, 
and the coordinates preserve the structure of the Rindler 
horizon for all $v$ in the sense that $g_{vv} = {\cal O}(r)$ and 
$g_{Av} = {\cal O}(r)$. $\Xi$~may hence be regarded as a 
Rindler version of a supertranslation vector field
\cite{Donn1,Donn2,Eling,Cai}, and it follows that the shell
imparts a Rindler supertranslation charge on the spacetime. 
The linearised stress-energy tensor vanishes 
for $v\ne v_0$ by construction, 
while at $v=v_0$ the stress-energy tensor is nonvanishing for generic~$f$, 
involving not just the Dirac delta but also the 
derivative of the Dirac delta~\cite{san}. 

It was shown in \cite{san} that the supertranslation charge imparted by 
the shell is detectable as a classical memory on a 
family of observers who prior to the shell are uniformly linearly accelerated, 
along the orbits of the boost Killing vector $\bar{\xi}$ \eqref{eq:boost-killing}. 
Assuming that each of these observers maintains their uniform 
linear acceleration on crossing the shell, 
as characterised at the shell by the appropriate local notion of 
acceleration in curved spacetime, 
the observers follow after the shell orbits of boost Killing 
vectors that differ from trajectory to trajectory, 
and the trajectory-dependence carries a memory of the planar 
inhomogeneity of the shell. In the rest of the paper we show that an 
accompanying memory exists also for a quantum field.

\section{Quantum memory for Rindler supertranslations}\label{BG}

We now turn to a real massless Klein-Gordon field
in the $r>0$ part of the shock wave spacetime~\eqref{eq:rindlershock1}. 
Geometrically, $r>0$ means that we only consider the right 
quadrant of the spacetime as shown in Figure~\ref{fig:diagram1}, 
but therein both the pre-shell region $v<v_0$ and the post-shell region $v>v_0$. 

\subsection{Classical field}

Working to linear order in~$f$, the Klein-Gordon field equation, 
$\nabla_a\nabla^a \phi =0$, takes the form 
\begin{align}
0 &= 2\kappa r \partial^2_r \phi 
+ 2 \partial_r \partial_v \phi 
+ \partial^2_x \phi 
+ \partial^2_y \phi 
+ 2 \kappa \partial_r \phi 
\notag 
\\
& \hspace{3ex}
- 4 rh (\partial^A f) \partial_r \partial_A \phi 
- 2 \kappa^{-1}rh (\partial^A \partial^B f ) \partial_A \partial_B \phi 
- 2h (\partial^A f) \partial_A \phi 
\nonumber 
\\
& \hspace{3ex}
- 2 \kappa^{-1}rh (\partial^A \partial_B \partial^B f ) \partial_A \phi 
+ \kappa^{-1} h ( \partial_B \partial^B f ) \partial_v \phi 
+  \kappa^{-1} r h' ( \partial_B \partial^B f ) \partial_r \phi
\ , 
\label{KGeqn}
\end{align}
where the derivatives in parentheses act only within the parentheses. 
We consider this equation first individually for $v<v_0$ and $v>v_0$, 
and then match the solutions at $v=v_0$. 

For $v<v_0$, the metric is given by~\eqref{metricRindler}, the 
terms proportional to $f$ in \eqref{KGeqn}
vanish, and \eqref{KGeqn} reduces to 
\begin{align}
0 &= 2\kappa r \partial^2_r \phi 
+ 2 \partial_r \partial_v \phi 
+ \partial^2_x \phi 
+ \partial^2_y \phi 
+ 2 \kappa \partial_r \phi 
\ . 
\label{rindler-KGeqn}
\end{align}
Separating \eqref{rindler-KGeqn} with the ansatz 
\begin{align}
\phi =  e^{-i \omega v} \phi_r(r) e^{i k_A x^A}
\ , 
\label{eq:phinought}
\end{align}
where $k_x, k_y, \omega \in \BbbR$, we find that $\phi_r(r)$ satisfies 
\begin{align}
2\kappa r \phi''_r + 2(\kappa - i \omega) \phi'_r - k^2 \phi_r = 0
\ , 
\label{kgsol0}
\end{align}
where $k = \sqrt{k_x^2 + k_y^2}$. Assuming $(k_x , k_y) \ne (0,0)$, so that $k>0$, 
\eqref{kgsol0} can be solved in terms of modified Bessel functions~\cite{dlmf}, 
and the solution that does not diverge at $r\to\infty$ is 
\begin{align}
\phi_r = N_1 
\bigg( \frac{2k \sqrt{r}}{\sqrt{2\kappa}}\bigg)^{i\omega /\kappa} 
K_{i \omega/\kappa}\bigg( \frac{2k \sqrt{r}}{\sqrt{2\kappa}}\bigg)
\ , 
\label{eq:phir-sol}
\end{align}
where $N_1$ is a normalisation factor. 
Solutions that are of positive frequency with respect to the boost Killing vector 
$\bar{\xi}$ \eqref{eq:boost-killing} are those with $\omega>0$. 

To fix the normalisation, we define the Klein-Gordon inner product 
on the null surfaces of constant~$v$, as in \cite{Gibbons:1975jb,Garriga:1990dp}. 
Using~\eqref{metricRindler}, 
the formula for the inner product becomes 
\begin{align}
\bigl\langle  W_1, W_2 \bigr\rangle =  
- i \int_0^\infty dr  \int_{\BbbR^2} dx\,dy 
\left( W_1 \partial_r W^*_2 -  W^*_2 \partial_r W_1 \right) 
\ . 
\label{eq:ip-def}
\end{align}
We have verified that this inner product is conserved, 
and it agrees with the inner product on surfaces that are deformed 
near $r=0$ to be spacelike and hit the Rindler 
horizon bifurcation point $(U,V) = (0,0)$. 
For a complete set of solutions that are positive 
frequency with respect to $\bar{\xi}$, we hence choose 
\begin{align}
\phi^0_{\omega, \bf{k}}
= 
\sqrt{\frac{\sinh{(\pi \omega / \kappa)}}{4 \pi^4 \kappa}}
\, 
e^{-i \omega v} 
\bigg( \frac{2k \sqrt{r}}{\sqrt{2\kappa}}\bigg)^{i\omega /\kappa} 
K_{i \omega/\kappa}\bigg( \frac{2k \sqrt{r}}{\sqrt{2\kappa}}\bigg)
e^{i k_A x^A}
\ , 
\label{eq:phi0-def}
\end{align}
where $\omega>0$ and ${\bf{k}} = (k_x, k_y) \in \BbbR^2 \setminus \{(0,0)\}$.  
The inner products are 
\begin{subequations}
\label{eq:phi0-orthonormality}
\begin{align}
\bigl\langle  \phi^0_{\omega, \bf{k}}, \phi^0_{\omega', \bf{k}'} \bigr\rangle 
&= \delta(\omega - \omega') \delta^2({\bf k} - {\bf k}') 
\ , 
\\ 
\bigl\langle  \phi^{0*}_{\omega, \bf{k}}, \phi^{0*}_{\omega', \bf{k}'} \bigr\rangle 
&= - \delta(\omega - \omega') \delta^2({\bf k} - {\bf k}') 
\ , 
\\
\bigl\langle  \phi^{0}_{\omega, \bf{k}}, \phi^{0*}_{\omega', \bf{k}'} \bigr\rangle 
&= 0 
\ . 
\end{align}
\end{subequations}

For $v > v_0$, we have $h(v-v_0) = 1$, 
and the terms proportional to 
$f$ in \eqref{KGeqn} do contribute. 
However, since the $v>v_0$ region of 
\eqref{eq:rindlershock1} is obtained from 
\eqref{metricRindler} by a diffeomorphism generated by the 
Rindler supertranslation 
vector field $\Xi$~\eqref{eq:Rindler-supertransfield}, 
and since we are working to linear order in~$f$, 
a complete set of mode solutions that are of positive 
frequency with respect to $\partial_v$ is 
\begin{align}
\phi^1_{\omega, \bf{k}}  &= \left( 1 - \Xi^a\partial_a \right) 
\phi^0_{\omega, \bf{k}}
\nonumber \\
&=  \left( 1 + i \kappa^{-1}\omega f + i \kappa^{-1} r k_A \partial^A f  \right) 
\phi^0_{\omega, \bf{k}}
\ , 
\label{eq:phi1-def}
\end{align}
where again $\omega>0$ and ${\bf{k}} = (k_x, k_y) \in \BbbR^2 \setminus \{(0,0)\}$.  
The Klein-Gordon inner product formula can be written down by applying 
the Rindler supertranslation diffeomorphism to~\eqref{eq:ip-def}, 
and the diffeomorphism construction guarantees that the inner products are 
\begin{subequations}
\begin{align}
\bigl\langle  \phi^1_{\omega, \bf{k}}, \phi^1_{\omega', \bf{k}'} \bigr\rangle 
&= \delta(\omega - \omega') \delta^2({\bf k} - {\bf k}') 
\ , 
\\ 
\bigl\langle  \phi^{1*}_{\omega, \bf{k}}, \phi^{1*}_{\omega', \bf{k}'} \bigr\rangle 
&= - \delta(\omega - \omega') \delta^2({\bf k} - {\bf k}') 
\ , 
\\
\bigl\langle  \phi^{1}_{\omega, \bf{k}}, \phi^{1*}_{\omega', \bf{k}'} \bigr\rangle 
&= 0 
\ . 
\end{align}
\end{subequations}

Now, consider the matching at $v=v_0$. 
We look for a solution to the 
linearised Klein-Gordon 
equation \eqref{KGeqn} as 
$\phi = \phi_0 + \phi_1 + {\mathcal O}(f^2)$, 
where $\phi_0$ has order $f^0$ and $\phi_1$ has order~$f$. 
Matching terms order by order shows that $\phi_0$ satisfies 
\eqref{rindler-KGeqn} and $\phi_1$ satisfies 
\begin{align}
0 &= 2\kappa r \partial^2_r \phi_1 
+ 2 \partial_r \partial_v \phi_1
+ \partial^2_x \phi_1
+ \partial^2_y \phi_1 
+ 2 \kappa \partial_r \phi_1
\notag 
\\
& \hspace{3ex}
- 4 rh (\partial^A f) \partial_r \partial_A \phi_0
- 2 \kappa^{-1}rh (\partial^A \partial^B f ) \partial_A \partial_B \phi_0
- 2h (\partial^A f) \partial_A \phi_0
\nonumber 
\\
& \hspace{3ex}
- 2 \kappa^{-1}rh (\partial^A \partial_B \partial^B f ) \partial_A \phi_0
+ \kappa^{-1} h ( \partial_B \partial^B f ) \partial_v \phi_0
+  \kappa^{-1} r h' ( \partial_B \partial^B f ) \partial_r \phi_0
\ . 
\label{KGeqn-order-f}
\end{align}
Assuming $\phi_0$ to be smooth across $v=v_0$, the terms involving 
$\partial_r\partial_v \phi_1$ and $h'$ in \eqref{KGeqn-order-f}
show that $\partial_r \phi_1$ has at $v=v_0$ a discontinuity,
and the matching condition reads 
\begin{align}
2 {[\partial_r \phi_1]}^{v_{0+}}_{v_{0-}} 
= -  \kappa^{-1} r ( \partial_B \partial^B f ) \partial_r \phi_0 |_{v_0} 
\ . 
\label{discontinuityeqn}
\end{align}
Assuming that $r \partial_r \phi_0$ is integrable at $r\to\infty$ and 
${[\phi_1]}^{v_{0+}}_{v_{0-}} \to 0$ as $r\to\infty$, 
which will hold for the functions below, we 
may integrate \eqref{discontinuityeqn} to 
\begin{align}
2 {[\phi_1]}^{v_{0+}}_{v_{0-}} 
= \kappa^{-1}( \partial_B \partial^B f ) \int_r^\infty dr \, r \partial_r \phi_0 |_{v_0}
\ . 
\label{discontinuityeqn-int}
\end{align}

Consider hence the solution $\widehat\phi_{\omega, \bf{k}}$ 
that is equal to $\phi^0_{\omega, \bf{k}}$ at $v<v_0$. 
Expanding this solution at $v>v_0$ in the basis 
$\left\{\phi^1_{\omega, \bf{k}}\right\}$, we write 
\begin{align}
\widehat\phi_{\omega, \bf{k}} = 
\begin{cases}
{\displaystyle \phi^0_{\omega, \bf{k}}} & \text{for $v<v_0$;}
\\[1ex]
{\displaystyle 
\int_0^\infty d\omega' \int d^2 {\bf k}' 
\left(  
\alpha_{\omega, {\bf k}; \omega', {\bf k}'} \phi^1_{\omega', \bf{k}'} 
+ \beta_{\omega, {\bf k}; \omega', {\bf k}'} \phi^{1*}_{\omega', \bf{k}'}  
\right)} 
& \text{for $v>v_0$,}
\end{cases}
\label{bgexpansion}
\end{align}
where the $\alpha$s and $\beta$s are the Bogoliubov coefficients between 
the $\left\{\phi^0_{\omega, \bf{k}}\right\}$ basis 
and the $\left\{\phi^1_{\omega, \bf{k}}\right\}$ basis~\cite{Birrell:1982ix}. 
Using \eqref{eq:phi1-def} and~\eqref{discontinuityeqn}, we find 
\begin{subequations}
\label{eq:Bogo-raw}
\begin{align}
\alpha_{\omega, {\bf k}; \omega', {\bf k}'} 
&= \delta(\omega - \omega') \delta^2({\bf k} - {\bf k}') 
+ \alpha^{(1)}_{\omega, {\bf k}; \omega', {\bf k}'}  + {\cal O}(f^2)
\ , 
\\
\beta_{\omega, {\bf k}; \omega', {\bf k}'}  
&= \beta^{(1)}_{\omega, {\bf k}; \omega', {\bf k}'} + {\cal O}(f^2)
\ , 
\end{align}
\end{subequations}
where the condition determining $\alpha^{(1)}$ and $\beta^{(1)}$ 
is that the equation  
\begin{align}
2 \int_0^\infty d\omega' \int d^2 {\bf k}' 
\left( 
\alpha^{(1)}_{\omega, {\bf k}; \omega', {\bf k}'} \phi^0_{\omega', {\bf k}'} 
+ \beta^{(1)}_{\omega, {\bf k}; \omega', {\bf k}'} \phi^{0*}_{\omega', {\bf k}'}  
\right)
&= 
\kappa^{-1}( \partial_B \partial^B f ) 
\int_r^\infty dr \, r \partial_r \phi^0_{\omega, {\bf k}} 
\notag
\\[1ex]
& \hspace{2.5ex}
+ 2 \,\Xi^a \partial_a \phi^0_{\omega, \bf{k}} 
\label{eq:alphabeta1-matching}
\end{align}
holds on the surface $v=v_0$. 
Evaluating $\left\langle \,\, \cdot \,\,, \phi^{0}_{\omega', {\bf k}'}\right\rangle_{\!v_0}$ 
and 
$\left\langle \,\, \cdot \,\,, \phi^{0*}_{\omega', {\bf k}'}\right\rangle_{\!v_0}$
on both sides of~\eqref{eq:alphabeta1-matching}, 
where $\left\langle \,\, \cdot \,\,, \,\, \cdot \,\,\right\rangle_{\!v_0}$ stands 
for the inner product \eqref{eq:ip-def} evaluated on the $v=v_0$ surface, 
and using \eqref{eq:phi0-orthonormality}, 
we hence obtain  
\begin{subequations}
\label{Bcoefficientsdef}
\begin{align}
\alpha^{(1)}_{\omega, {\bf k}; \omega', {\bf k}'} 
&= 
\bigl\langle 
\Xi^a \partial_a \phi^0_{\omega, {\bf k}} , \phi^{0}_{\omega', {\bf k}'} 
\bigr\rangle_{\!v_0}
+ {(2\kappa)}^{-1}
\left\langle 
(\partial_B \partial^B f) \int_r^\infty dr \, r \partial_r \phi^0_{\omega, {\bf k}} , \, 
\phi^{0}_{\omega', {\bf k}'} 
\right\rangle_{\!\!v_0}
\ , 
\label{Bcoefficientsdef-alpha}
\\
\beta^{(1)}_{\omega, {\bf k}; \omega', {\bf k}'} 
&= 
- \bigl\langle 
\Xi^a \partial_a \phi^0_{\omega, {\bf k}} , \phi^{0*}_{\omega', {\bf k}'} 
\bigr\rangle_{\!v_0} 
- {(2\kappa)}^{-1}
\left\langle
(\partial_B \partial^B f) \int_r^\infty dr \, r \partial_r \phi^0_{\omega, {\bf k}} , \, 
\phi^{0*}_{\omega', {\bf k}'} 
\right\rangle_{\!\!v_0}
\ . 
\label{Bcoefficientsdef-beta}
\end{align}
\end{subequations}
Writing 
\begin{align}
\Xi^a\partial_a
\phi^0_{\omega, \bf{k}}
= - i \kappa^{-1}\omega f 
\phi^0_{\omega, \bf{k}}
- i \kappa^{-1} r k_A \partial^A f 
\phi^0_{\omega, \bf{k}}
\ , 
\label{eq:Xi-on-phi}
\end{align}
we can evaluate \eqref{Bcoefficientsdef} using 
formula 6.576.4 in \cite{Grad-Ryz} 
and the integral identity that we give in Appendix~\ref{app:K-Bessel-integral}. 
We find 
\begin{subequations}
\label{eq:alpha1beta1-cont}
\begin{align}
\alpha^{(1)}_{\omega, {\bf k}; \omega', {\bf k}'} 
& = - \frac{i \omega \widetilde{f}(\tilde {\bf k})}{4 \pi^2 \kappa}
\left(\frac{k}{k'}\right)^{ \! i\omega/\kappa}  
 \left[1 + \frac{i\omega}{2 \kappa}\left(1- \frac{k^{2}}{k^{\prime 2}} \right)
{}_2F_1 \left(1+\frac{i\omega}{\kappa}, 1; 2; 1- \frac{k^{2}}{k^{\prime 2}} \right)
\right]
\notag
\\[1ex]
&\hspace{5ex}
\times
\delta(\omega-\omega')
\notag
\\[1ex]
&\hspace{3ex} 
- \frac{i \omega \widetilde{f}(\tilde {\bf k}) \sqrt{\sinh(\pi \omega / \kappa) \sinh(\pi \omega' / \kappa)}}{32 \pi^2 \kappa^3 \sinh\bigl(\pi(\omega+\omega')/(2\kappa)\bigr)}
\, 
P\left(\frac{\omega + \omega'}{\sinh\bigl(\pi(\omega-\omega')/(2\kappa)\bigr)}\right) 
\notag
\\
&\hspace{6ex}
\times 
\left( \frac{k}{k^{\prime}} \right)^{ \! i\omega'/\kappa} 
\left(1- \frac{k^{2}}{k^{\prime 2}} \right) 
{}_2F_1 \! \left(
1+\frac{i(\omega+\omega')}{2 \kappa}, 1 + \frac{i(\omega'-\omega)}{2 \kappa}; 2; 1- \frac{k^{2}}{k^{\prime 2}} 
\right)
\notag
\\[1ex]
&\hspace{3ex}
- \frac{i\widetilde{f}(\tilde {\bf k})
e^{-i(\omega - \omega') v_0} \sqrt{\sinh(\pi \omega / \kappa) \sinh(\pi \omega' / \kappa)}}{8 \pi^4 \kappa} 
{\left(\frac{k}{\sqrt{2} \kappa}\right)}^{-i \omega/\kappa}  
{\left(\frac{k'}{\sqrt{2} \kappa}\right)}^{i \omega'/\kappa} 
\notag \\
&\hspace{5ex}
\times  \Gamma \left(1 + \frac{i \omega}{\kappa}\right)\, 
\Gamma \left(1 - \frac{i \omega'}{\kappa}\right) \, 
\Gamma \left(1 + \frac{i \omega}{\kappa} - \frac{i \omega'}{\kappa} \right) 
\notag \\
&\hspace{5ex} 
\times \Biggl[
\frac{{\tilde{k}}^2}{k^2} \left(1 - \frac{i \omega'}{\kappa}\right) 
{}_2F_1\left( 2, 1 + \frac{i \omega}{\kappa}; 3 + \frac{i \omega}{\kappa} - \frac{i \omega'}{\kappa} ; 
1 - \left( \frac{k'}{k}\right)^2 \right) 
\notag \\
&\hspace{9ex} 
+ \frac{4 {k_A} {\tilde{k}}^A}{{(k')}^2} 
\left(1 + \frac{i \omega}{\kappa}\right) 
{}_2F_1\left( 2, 1 - \frac{i \omega'}{\kappa}; 3 + \frac{i \omega}{\kappa} - \frac{i \omega'}{\kappa} ; 
1 - \left( \frac{k}{k'}\right)^2 \right) 
\Biggr]
\ , 
\label{alpha}
\end{align}
\begin{align}
\beta^{(1)}_{\omega, {\bf k}; \omega', {\bf k}'} 
&=
\frac{i \omega \widetilde{f}(\tilde {\bf k}_+) \sqrt{\sinh(\pi \omega / \kappa) \sinh(\pi \omega' / \kappa)}}{32 \pi^2 \kappa^3 \sinh\bigl(\pi(\omega+\omega')/(2\kappa)\bigr)}
\left(\frac{\omega - \omega'}{\sinh\bigl(\pi(\omega-\omega')/(2\kappa)\bigr)}\right)
\notag
\\[1ex]
&\hspace{4ex}
\times 
\left( \frac{k}{k'} \right)^{ \! i\omega'/\kappa}
\left(1- \frac{k^{2}}{k^{\prime 2}} \right) 
{}_2F_1 \! \left(1+\frac{i(\omega+\omega')}{2 \kappa}, 1 + \frac{i(\omega'-\omega)}{2 \kappa}; 2; 1- \frac{k^{2}}{k^{\prime 2}} \right)
\notag
\\[1ex]
&\hspace{3ex}
+ \frac{i\widetilde{f}(\tilde {\bf k}_+)
e^{-i(\omega + \omega') v_0} \sqrt{\sinh(\pi \omega / \kappa) \sinh(\pi \omega' / \kappa)}}{8 \pi^4 \kappa} 
{\left(\frac{k}{\sqrt{2} \kappa}\right)}^{-i \omega/\kappa}  
{\left(\frac{k'}{\sqrt{2} \kappa}\right)}^{-i \omega'/\kappa} 
\notag \\
&\hspace{5ex}
\times  \Gamma \left(1 + \frac{i \omega}{\kappa}\right)\, 
\Gamma \left(1 + \frac{i \omega'}{\kappa}\right) \, 
\Gamma \left(1 + \frac{i \omega}{\kappa} + \frac{i \omega'}{\kappa} \right) 
\notag \\
&\hspace{5ex} 
\times \Biggl[
\frac{{\tilde{k}}_+^2}{k^2} \left(1 + \frac{i \omega'}{\kappa}\right) 
{}_2F_1\left( 2, 1 + \frac{i \omega}{\kappa}; 3 + \frac{i \omega}{\kappa} + \frac{i \omega'}{\kappa} ; 
1 - \left( \frac{k'}{k}\right)^2 \right) 
\nonumber \\
&\hspace{9ex} 
+\frac{4 {k_A} {\tilde{k}}_+^A}{{(k')}^2} \left(1 + \frac{i \omega}{\kappa}\right) 
{}_2F_1\left( 2, 1 + \frac{i \omega'}{\kappa}; 3 + \frac{i \omega}{\kappa} + \frac{i \omega'}{\kappa}; 
1 - \left( \frac{k}{k'}\right)^2 \right) 
\Biggr]
\ , 
\label{beta}
\end{align}
\end{subequations}
where $\tilde {\bf k} = {\bf k} - {\bf k}'$, $\tilde {\bf k}_+ = {\bf k} + {\bf k}'$, 
$\widetilde f$ is the Fourier transform of $f$ as defined by  
\begin{align}
{\widetilde f}({\bf k}) = \int e^{i k_A x^A} f(x,y) \, dx\,dy
\ ,  
\end{align}
and $P$ stands for the Cauchy principal value. 

We see that both 
$\alpha^{(1)}_{\omega, {\bf k}; \omega', {\bf k}'}$
and 
$\beta^{(1)}_{\omega, {\bf k}; \omega', {\bf k}'}$
are nonvanishing for generic~$f$. 
$\alpha^{(1)}_{\omega, {\bf k}; \omega', {\bf k}'}$ 
is distributional at $\omega=\omega'$, 
having both a Dirac delta and a Cauchy principal value there, 
whereas $\beta^{(1)}_{\omega, {\bf k}; \omega', {\bf k}'}$ 
has no distributional singularities.

\subsection{Quantised field}

We are now ready to read off the quantum memory associated with the shell. 

Since the modes $\phi^0_{\omega, \bf{k}}$ 
are of positive frequency with respect to the 
Killing vector $\partial_v$ for $v<v_0$, 
and the modes $\phi^1_{\omega, \bf{k}}$ are of positive 
frequency with respect to the Killing vector $\partial_v$ for $v>v_0$, 
we can quantise the field in each region by adopting these 
modes as the positive frequency basis functions. 
As $\partial_v$ generates a pure boost in each region, 
the Fock vacua that ensue are of the Rindler type, 
seen as a no-particle state by the uniformly accelerated 
observers who follow the orbits of~$\partial_v$. 

However, the Bogoliubov transformation \eqref{bgexpansion} 
between the two sets of modes is 
nontrivial, and in particular it involves nonvanishing beta-coefficients. 
It follows that the 
two Rindler vacua are not equivalent: 
if the field is initially prepared in the $v<v_0$ Rindler vacuum, 
the field is no longer in the Rindler vacuum for $v>v_0$. 

Hence, the shell creates Rindler particles that 
contain information about the classical supertranslation 
field~$\Xi$, and specifically
about the planar profile $f$ of the supertranslational shockwave.
This is a quantum counterpart of the classical Rindler supertranslation 
memory found in~\cite{san}. 

In particular, if the field is prepared in the 
Minkowski vacuum at $v < v_0$, 
the reduced density matrix in the right Rindler wedge 
will acquire non-thermal corrections for $v > v_0$. 
The extra Rindler particles created by the shell 
change the entanglement between observers who reside in the opposite Rindler wedges. 
We shall analyse this phenomenon in the next section.

\section{Entanglement due to Rindler supertranslations}\label{entanglement}

\subsection{The entanglement setup}

It is well appreciated that Minkowski vacuum contains nonlocal 
spatial correlations that can be harvested by localised quantum 
systems~\cite{valentini-corr,Reznik:2002fz,Reznik:2003mnx,Pozas-Kerstjens:2015gta}. 
For a pair of localised observers who follow the orbits 
of a boost Killing vector, accelerating in opposite 
directions with acceleration of magnitude~$a$, these 
quantum correlations appear as a two-mode squeezed 
state, and each of the individual observers experiences 
the state as thermal in the Unruh temperature $a/(2\pi)$~\cite{unruh}. 

Suppose now that one of the accelerated observers 
goes through the shock wave~\eqref{eq:rindlershock1}. 
How does the shock wave affect the quantum correlations between the two observers? 

\begin{figure}[t]
\includegraphics[width=\linewidth]{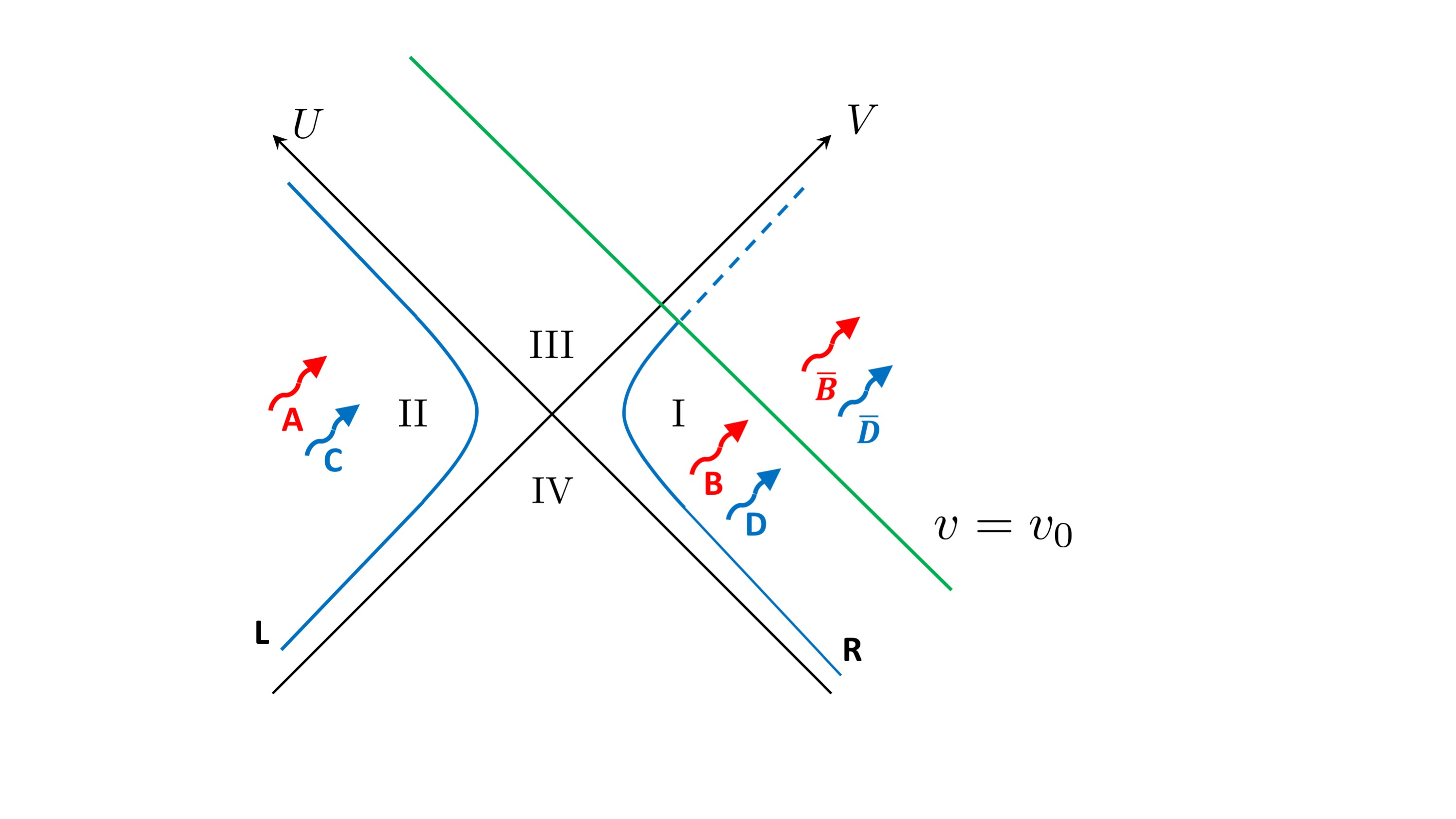} 
\caption{A pair of uniformly accelerated observers 
in the shock wave spacetime of Figure~\ref{fig:diagram1}. 
Luke (L)
accelerates leftward in quadrant~II\null. 
Rob (R)
accelerates rightward in quadrant~I, crossing the wave. 
After crossing the wave, Rob's trajectory is no 
longer a pure hyperbola in the two dimensions shown in the diagram, 
but it involves a perturbative correction due to the wave~\cite{san}. 
Luke couples to field modes labelled $A$ and~$C$. 
Rob couples before the wave-crossing to field modes labelled $B$ and~$D$, 
and after the wave-crossing to field modes labelled ${\bar B}$ and~${\bar D}$. 
\label{fig:luke-and-rob}}
\end{figure}

To set up the notation, we call the two observers respectively Luke and Rob, 
with Luke accelerating to the left and Rob accelerating to the right, 
as shown in Figure~\ref{fig:luke-and-rob}. 
In the past of the shell, the Minkowski vacuum $| 0\rangle_M$ 
can be written as \cite{Birrell:1982ix} 
\begin{align}
| 0\rangle_M = 
\prod_{\omega, {\bf k}} 
\sqrt{1 - e^{- 2 \pi \omega/\kappa}} \; \sum_{n=0}^\infty 
e^{-n \pi \omega/\kappa} \;\; | n\rangle_{L,\omega, {\bf k}}  \otimes | n\rangle_{R,\omega, {\bf k}}
\ , 
\label{minvacuum}
\end{align}
where $| n\rangle_{R,\omega, {\bf k}}$ are the Fock basis states 
in region I in the notation of Section~\ref{BG} and 
$| n\rangle_{L,\omega, {\bf k}}$ are the corresponding 
Fock basis states in region~II\null. 
If the observers' proper acceleration has magnitude~$a$, 
the frequency with respect to the observers' proper time is 
related to $\omega$ by by~$(a/\kappa) \omega$. 

To describe the correlations between Luke and Rob after Rob has crossed the shock wave, 
we need to write \eqref{minvacuum} in Rob's new basis, obtained from the old basis 
by the Bogoliubov transformation~\eqref{eq:Bogo-raw}. 

We simplify this problem in two ways. First, 
instead of the continuous labels $\omega$ and~${\bf k}$, 
we postulate that Rob and Luke each couple to just two modes of the field. 
This sidesteps the technical issue that the product over the modes on the 
right-hand side of \eqref{minvacuum} is not mathematically well defined, 
and the related open questions of quantifying entanglement 
with continuously-labelled mode sets. 
Conditions under which this postulate may provide a reasonable 
approximation in a sense of wave packets 
are discussed in~\cite{Bruschi:2010mc}. 
Second, we truncate the initial state of each mode to keep just the 
$n=0$ and $n=1$ states. From \eqref{minvacuum} we see that is 
a good approximation for the high energy modes, $\omega/\kappa\gg1$.

\subsection{Before the wave}

We denote the two modes to which Luke couples by $A$ and~$C$, 
and the two modes to which Rob couples by $B$ and~$D$. 
Before the shock wave, we take the state to be 
\begin{align}
|\Phi \rangle = |\phi_{1} \rangle \otimes |\phi_{2}  \rangle
\ , 
\label{statedef}
\end{align}
where 
\begin{subequations}
\label{statedef2}
\begin{align}
|\phi_{1} \rangle  
&= 
\frac{1}{\sqrt{1+p^2}}\left( |0\rangle_A \otimes |0 \rangle_B 
+ p |1\rangle_A \otimes |1 \rangle_B  \right) 
\ , 
\label{statedef2-one}
\\
|\phi_{2} \rangle  
&= 
\frac{1}{\sqrt{1+q^2}}\left( |0\rangle_C \otimes |0 \rangle_D 
+ q |1\rangle_C \otimes |1 \rangle_D  \right)
\ , 
\end{align}
\end{subequations}
and $p$ and $q$ are real-valued parameters.  
$|\Phi \rangle$ is a good approximation to the high-frequency regime in 
\eqref{minvacuum} when 
$0 < p \ll 1$ and $0 < q \ll 1$, 
but in what follows we consider the more general situation 
in which $p$ and $q$ are allowed to be arbitrary. 

We quantify the entanglement in $|\Phi \rangle$ by the negativity~${\mathcal N}$, 
reviewed in Appendix~\ref{app:negativity}. 
$|\phi_{1} \rangle$ is bipartite in $A\leftrightarrow B$ and has negativity $p/(1+p^2)$; 
similarly, $|\phi_{2} \rangle$ is bipartite in $C\leftrightarrow D$ and has negativity 
$q/(1+q^2)$. 
There is clearly no entanglement in the subsystems $A\leftrightarrow C$, $A\leftrightarrow D$, 
$B\leftrightarrow D$ and $B\leftrightarrow C$, and the corresponding negativities vanish. Collecting, 
the nonvanishing negativities are 
\begin{subequations}
\begin{align}
{\mathcal N}_{A\leftrightarrow B} &= \frac{p}{1+p^2}
\ , 
\label{eq:NAB-initial}
\\
{\mathcal N}_{C\leftrightarrow D} &= \frac{q}{1+q^2}
\ .  
\end{align}
\end{subequations}
The total Rob-Luke negativity is  
${\mathcal N}_{A\leftrightarrow B} + {\mathcal N}_{C\leftrightarrow D} = p/(1+p^2) + q/(1+q^2)$.

\subsection{After the wave}

After Rob has crossed the wave, we denote the two modes to which Rob couples by 
$\bar B$ and~$\bar D$. We 
write the Bogoliubov coefficients from $\left\{B,D\right\}$ to $\left\{\bar B, \bar D\right\}$ as in 
\eqref{bgexpansion} and~\eqref{eq:Bogo-raw}, 
\begin{subequations}
\label{eq:22-bogos}
\begin{align}
\alpha_{k ,k^\prime} &= 
\delta_{k,k^\prime} 
+ \lambda \; \alpha^{(1)}_{k ,k^\prime} + \lambda^2 \alpha^{(2)}_{k ,k^\prime} 
+ {\mathcal O}\bigl(\lambda^3\bigr)
\ ,
\\
\beta_{k ,k^\prime} &= 
\lambda \; \beta^{(1)}_{k ,k^\prime} 
+ \lambda^2  \beta^{(2)}_{k ,k^\prime}
+ {\mathcal O}\bigl(\lambda^3\bigr)
\ , 
\end{align}
\end{subequations}
where $k \in \left\{ B,D \right\}$ and $k' \in \left\{ \bar B, \bar D \right\}$, 
and we have included a formal perturbative parameter 
$\lambda$ to facilitate the book-keeping in the perturbative expansion. 

For \eqref{eq:22-bogos} to provide a mathematically 
consistent Bogoliubov transformation, $\alpha_{k ,k^\prime}$ and 
$\beta_{k ,k^\prime}$ must satisfy the Bogoliubov identities~\cite{Birrell:1982ix}, 
which imply in the linear order that $\alpha^{(1)}$ is anti-Hermitian and $\beta^{(1)}$ 
is symmetric, while in higher orders they imply relations 
involving the higher-order coefficients 
(see Appendix A of~\cite{Doukas:2014bja}). 
Reducing \eqref{Bcoefficientsdef} to \eqref{eq:22-bogos} 
in a way that satisfies these identities would need additional 
input about the reduction, such as 
a construction of suitable wave packets~\cite{Bruschi:2010mc}, 
and we shall not attempt to provide this input here. 
Instead, we shall proceed without specifying the explicit form of 
$\alpha^{(1)}_{k ,k^\prime}$ and $\beta^{(1)}_{k ,k^\prime}$. 
This will suffice to demonstrate that the shock wave does 
change the bipartite entanglements in the system. 

\subsection{Changes in entanglement}

In terms of the Bogoliubov transformation~\eqref{eq:22-bogos}, 
we have \cite{Birrell:1982ix,fabbri}
\begin{subequations}
\label{eq:discr-bogo-f}
\begin{align}
a_k^\dagger &= \sum_{k'} \left( \alpha_{k,k'} {\bar a}_{k'}^\dagger - \beta_{k,k'} {\bar a}_{k'} \right) 
\ , 
\\ 
| 0 \rangle_B | 0 \rangle_D  &= N e^{\frac{1}{2} \sum_{mn} V_{mn} {\bar a}^\dagger_m {\bar a}^\dagger_n}
| 0 \rangle_{\bar B} | 0 \rangle_{\bar D}
\ , 
\label{vacuum}
\end{align}
\end{subequations}
where 
$a_m$ and $a^\dagger_m$ are the annihilation and creation 
operators for Rob's early time modes $B$ and~$D$, 
${\bar a}_m$ and ${\bar a}^\dagger_m$ are the annihilation and creation 
operators for Rob's late time modes ${\bar B}$ and~${\bar D}$, 
$V_{mn} = \beta^\dagger_{mp} \bigl(\alpha^{-1}\bigr)^\dagger_{pn}$, 
and $N$ is a normalisation constant. 
Using~\eqref{eq:discr-bogo-f}, 
we can transform the state $|\Phi\rangle$ \eqref{statedef} 
to the late time basis and analyse the negativity for the 
bipartite subsystems of interest. 
We shall omit the calculational details and just describe the outcome. 

\subsubsection{Subsystem $A$ and ${\bar B}$.}

Consider the system formed by $A$ and~${\bar B}$. 
Before the wave this was the single pair $|\phi_{1} \rangle$~\eqref{statedef2-one}, 
one mode coupling to Luke and the other to Rob, with negativity 
${\mathcal N}_{A\leftrightarrow B}$~\eqref{eq:NAB-initial}. 

After the wave, 
the reduced density matrix $\rho_{A\leftrightarrow {\bar B}}$ 
is obtained by tracing out $C$ and~${\bar D}$. Keeping terms of order~$\lambda^2$, 
we find that the partial transpose 
$\rho^T_{A\leftrightarrow {\bar B}}$ is a 12 by 12 matrix, 
and the correction to ${\mathcal N}_{A\leftrightarrow B}$ \eqref{eq:NAB-initial} 
starts in order~$\lambda^2$. 
We consider this correction here in the limit in which the 
diagonal elements of $\alpha^{(1)}$ and $\beta^{(1)}$ are negligible 
compared with the off-diagonal elements; 
this limit 
can be motivated by observing that in the continuous label case~\eqref{eq:alpha1beta1-cont}, 
the last two terms in \eqref{alpha} 
and the first term in \eqref{beta} vanish on the diagonal. 
The correction to the negativity comes then entirely 
from the correction to the single negative eigenvalue of $\rho^T_{A\leftrightarrow B}$, 
and we find 
\begin{align}
{\mathcal N}_{A \leftrightarrow {\bar B}} 
&=  
\frac{p}{1+p^2} 
- \lambda^2 \Biggl( \frac{p^2(1+2q^2) + q^2 + p(1 + 5q^2)}{2(1+p^2)(1+q^2)} 
\;  \bigl|\alpha^{(1)}_{B,D}\bigr|^2 
\notag 
\\
& \hspace{17ex}
+ \frac{ p\left[ p^2q^2 + 1 + 2q^2 + p(3 + 5q^2)\right] \left( 1+ q^2 \right) - 2p^4 q^4}{2p(1+p^2){(1+q^2)}^2} 
\;  \bigl|\beta^{(1)}_{B,D}\bigr|^2 \Biggr)
\notag 
\\
& \hspace{11ex}
+ {\mathcal O}(\lambda^3)
\ . 
\label{nab}
\end{align}

As both $\alpha^{(1)}$ and $\beta^{(1)}$ appear in~\eqref{nab}, 
the change in the entanglement is due in part to particle 
creation and in part to mode mixing. 
The sign of the correction term
in \eqref{nab} is typically negative, that is, entanglement is degraded. 
However, it can be arranged to be positive if
$\beta^{(1)}_{B,D}$ is nonzero  
and $p$ and $q$ are sufficiently large. 
An increase in the entanglement, when it occurs, 
is hence necessarily associated with particle creation. 

For the Minkowski vacuum state \eqref{minvacuum} in the high frequency limit,  
$\omega/ \kappa \gg 1$, the correction term 
in \eqref{nab}
is negative since in this case $0<p\ll1$ and $0<q\ll1$. 
The wave has hence degraded the entanglement between $A$ and~${\bar B}$.

An interesting special case occurs when $q=0$ and $p=1$: 
there is then initially only one entangled pair, and this pair is  
prepared in the maximally entangled Bell state. 
The initially maximal entanglement is degraded, 
as seen from the sign of the correction in~\eqref{nab}. 
This system is mathematically identical to the cavity 
system considered in~\cite{voyage}, and \eqref{nab} 
agrees with the correction found therein.

\subsubsection{Subsystem ${\bar B}$ and ${\bar D}$}

Consider the system formed by ${\bar B}$ and ${\bar D}$. 
Before the wave this was a system of two completely unentangled 
modes coupled to Rob, with vanishing negativity. 

After the wave, 
the reduced density matrix $\rho_{{\bar B} \leftrightarrow {\bar D}}$ 
is obtained by tracing out Luke's modes $A$ and~$C$. 
Keeping terms of order~$\lambda^2$, 
we find that the partial transpose 
$\rho^T_{{\bar B} \leftrightarrow {\bar D}}$ is a 14 by 14 matrix. 
The leading correction to the negativity appears in order~$\lambda$, and we find 
\begin{align}
{\mathcal N}_{{\bar B} \leftrightarrow {\bar D}} 
&= 
\lambda \; \frac{2p^2q^2 \bigl|\beta^{(1)}_{B,D}\bigr|}{(1+p^2)(1+q^2)} 
+ {\mathcal O}(\lambda^2)
\ . 
\label{nbd}
\end{align}
The wave has hence entangled Rob's two modes. 
As the Bogoliubov coefficient entering 
\eqref{nbd} is~$\beta^{(1)}$,  
the leading order entanglement creation is due to particle creation, 
not due to mode mixing.

\subsubsection{Subsystem ${\bar B}$ and $C$}

Consider finally the system formed by ${\bar B}$ and~$C$. 
Before the wave this was a system of two completely unentangled 
modes, one coupled to Luke and the other to Rob, with vanishing negativity. 

After the wave, 
the reduced density matrix $\rho_{{\bar B} \leftrightarrow C}$ 
is obtained by tracing out Luke's mode $A$ and Rob's mode~${\bar D}$. 
Keeping terms of order~$\lambda^2$, 
we find that the partial transpose 
$\rho^T_{{\bar B} \leftrightarrow C}$ is a 12 by 12 matrix, 
and the first contribution to 
${\mathcal N}_{{\bar B}\leftrightarrow C}$ comes in order~$\lambda^2$. 
Specialising again to the limit in which the 
diagonal elements of $\alpha^{(1)}$ and $\beta^{(1)}$ are negligible 
compared with the off-diagonal elements, we find 
\begin{align}
{\mathcal N}_{{\bar B} \leftrightarrow C} 
&= 
\lambda^2 
\Biggl( 2 q^2 \max \! 
\left(
2 p^2 q^2 \bigl|\alpha^{(1)}_{B,D}\bigr|^2 - \bigl|\beta^{(1)}_{B,C}\bigr|^2  , 0
\right)
\notag
\\
& \hspace{8ex}
+ p^2 \max \! 
\left( 
(2q^2-1) \bigl|\alpha^{(1)}_{B,D}\bigr|^2 + 2 (q^4-1) \bigl|\beta^{(1)}_{B,C}\bigr|^2 , 0 
\right) 
\Biggr)
+ {\mathcal O}(\lambda^3)
\ . 
\label{nbc}
\end{align}
There exist parameter ranges in which ${\mathcal N}_{{\bar B} \leftrightarrow C}>0$, 
and the entanglement generation comes from a mixture of particle creation and mode mixing effects.

\subsubsection{Quantum monogamy and negativity}

A curious property in the above negativity results 
is that entanglement generation can in certain circumstances 
happen already in order~$\lambda$, as seen in~\eqref{nbd}, but entanglement 
degradation will happen only in order~$\lambda^2$, as seen in~\eqref{nab}.
For example, suppose that $p=1$, and consider the 
entanglement of $\bar B$ with $A$ and with~$\bar D$. 
${\mathcal N}_{A \leftrightarrow {\bar B}}$ \eqref{nab}
has decreased from the maximal entanglement value $1/2$ in order~$\lambda^2$, but 
${\mathcal N}_{{\bar B} \leftrightarrow {\bar D}}$ \eqref{nbd}
has increased from the vanishing entanglement value $0$ already in order~$\lambda$. 
This might at first sight appear to be at tension with the monogamy of entanglement, 
which states that given a pair of maximally entangled systems, 
neither member of the pair can be entangled with a third system~\cite{CKW-ineq}.

However, there is in fact no tension. 
The reason is that the monogamy inequality that relates to negativity is not linear but quadratic, 
taking in the present situation the 
form \cite{karmarkar} 
(for related discussion see \cite{regula1,Choi,regula2})
\begin{align}
{\mathcal N}^2_{{\bar B} \leftrightarrow AC{\bar D}}  
\ge 
{\mathcal N}^2_{A \leftrightarrow {\bar B}} 
+ {\mathcal N}^2_{{\bar B} \leftrightarrow C} 
+ {\mathcal N}^2_{{\bar B} \leftrightarrow {\bar D}}
\ . 
\label{monogamy}
\end{align}
An explicit calculation of the entanglement of ${\bar B}$ with $A$, $C$ and ${\bar D}$ 
(negative eigenvalues of the $16 \times 16$ matrix obtained by taking the partial transpose of late time $\rho = |\Phi \rangle \langle \Phi |$ with respect to ${\bar B}$) 
shows that ${\mathcal N}_{{\bar B} \leftrightarrow AC{\bar D}}$ does not obtain a correction at order $\lambda$ from the maximal entanglement value of~$1/2$, for the maximal case when $p = 1$ and $q =1$, but gets possible corrections starting from order~$\lambda^2$. Hence, none of the terms on either side of the inequality \eqref{monogamy} are linear in~$\lambda$, since the only linear order term generated in ${\mathcal N}_{{\bar B} \leftrightarrow {\bar D}}$ in \eqref{nbd} becomes order $\lambda^2$ upon squaring 
on the right hand side\null. Inequality \eqref{monogamy} is thus satisfied to order $\lambda$ and there is no contradiction. 

From our results in \eqref{nab}, \eqref{nbd} and \eqref{nbc}, it
is straightforward to check that the 
$\lambda^2$ correction term on the right hand side of \eqref{monogamy} 
for the maximal case when $p = 1$ and $q =1$ is given by
\begin{equation}
{\mathcal N}^2_{A \leftrightarrow {\bar B}} 
+ {\mathcal N}^2_{{\bar B} \leftrightarrow C} 
+ {\mathcal N}^2_{{\bar B} \leftrightarrow {\bar D}} 
= \frac{1}{4} - \frac{5}{4} \left( \bigl|\alpha^{(1)}_{B,D}\bigr|^2 + 2 \bigl|\beta^{(1)}_{B,D}\bigr|^2  \right) 
\lambda^2
+ {\mathcal O}(\lambda^3)
\ . 
\label{eq:}
\end{equation}
As is expected, the coefficient of the $\lambda^2$ term is nonpositive. For the left hand side of \eqref{monogamy}, to obtain the explicit form of the $\lambda^2$ term in ${\mathcal N}^2_{{\bar B} \leftrightarrow AC{\bar D}}$, one arrives at a $81 \times 81$ partially transposed matrix with respect ${\bar B}$. Calculating the eigenvalues of such a large matrix is highly non-trivial computationally and beyond the scope of the present work. However, we refer the reader to \cite{karmarkar, regula1,Choi,regula2} wherein a detailed discussion of the above monogamy inequality for negativity can be found. 

\section{Discussion}\label{discsection}

We have shown that a classical supertranslation hair implanted on a Rindler horizon by a shock wave induces in quantum field theory a quantum supertranslation memory that modulates the entanglement between the two opposing Rindler wedges. In the Bogoliubov coefficient description, this memory involves nontrivial alpha-coefficients and nontrivial beta-coefficients, so that there is both particle creation and mode mixing. Within an entanglement analysis that truncates the number of field modes, we identified subsystems whose entanglement is degraded and subsystems whose entanglement is enhanced, and the entanglement effect appears to be robust against the input used in the truncation. Similar entanglement degradation and generation has been previously found in cavity systems in non-inertial 
motion~\cite{voyage,motion,bruschi}.

\begin{figure}
\includegraphics[width=\linewidth]{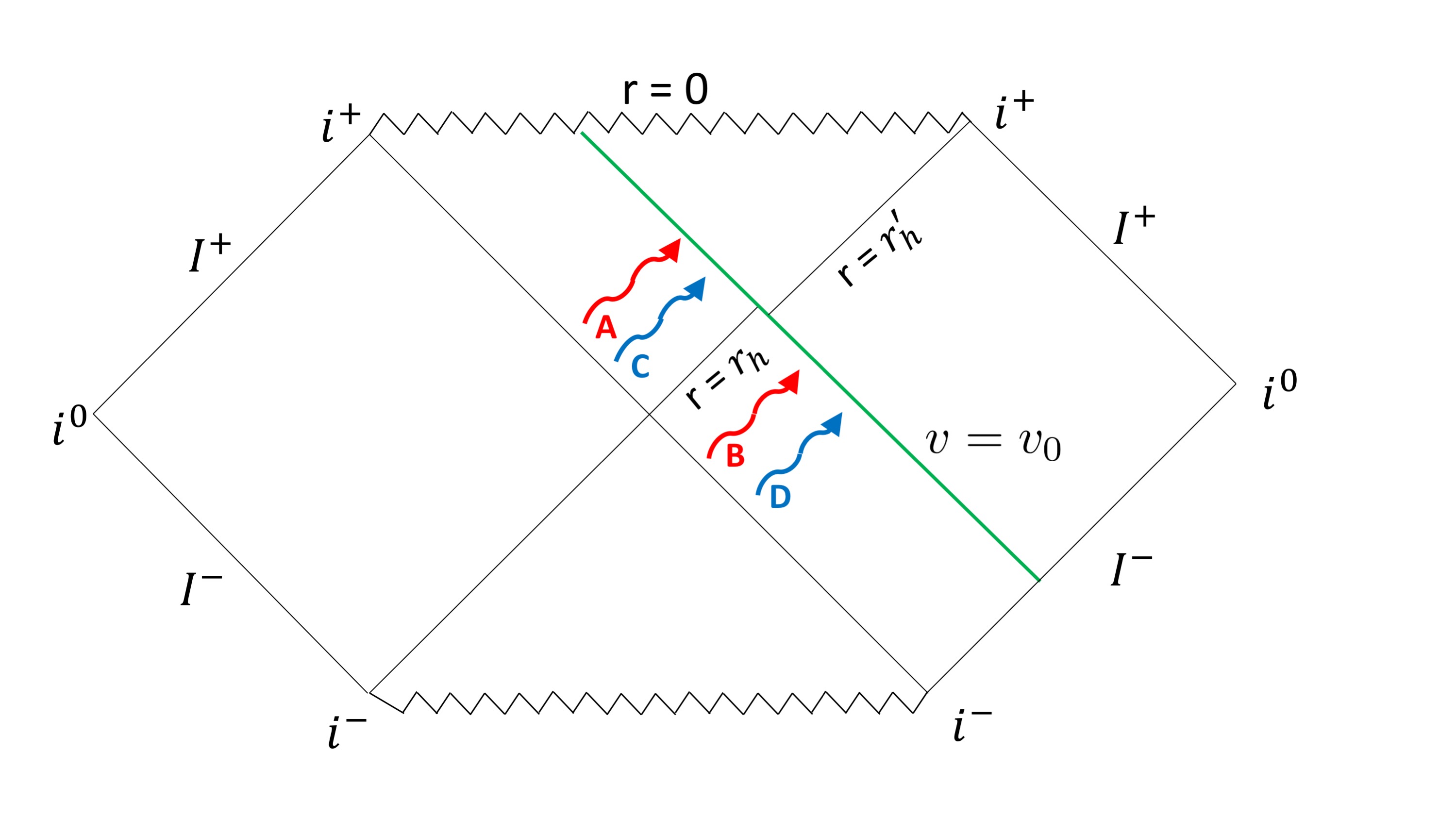}
\caption{The extended Schwarzschild spacetime with an infalling, 
supertranslation-implanting shock wave at $v=v_0$~\cite{HPS2}. 
Two Hawking pairs are shown, 
$A \leftrightarrow B$ (red) and $C \leftrightarrow D$ (blue).
\label{fig:schsup}}
\end{figure}

The linearised stress-energy tensor of the
supertranslated metric \eqref{eq:rindlershock1}
is given in equation (2.9) of~\cite{san}.
This stress-energy tensor is linear in the function $f$
that characterises the planar asymmetry of the shock wave,
and for a generic $f$ the stress-energy tensor
breaks the usual energy conditions somewhere,
in particular on crossing the Rindler horizon
from region I to region III in Figure~\ref{fig:diagram1}. 
Further, for a given~$f$, these violations become arbitrarily 
large near~${\mathcal{I}}^-$: this phenomenon stems from the 
diverging norm of the Rindler Killing vector $\partial_v$ 
near the infinity, and indicates that linearised perturbation
 theory is not reliable to arbitrarily large~$r$.
However, it is possible to amend the shock wave by adding to $T_{vv}$ a
uniform surface energy density~$\mu$, as shown in equation (2.10)
of~\cite{san}, and for a wave released at any finite value of~$r$.
The amended stress-energy tensor may still break the energy condition
due to quantum effects, however we note that the time averaged quantity,
 $\int_{v_0 - \epsilon}^{v_0 + \epsilon}dv T_{vv} $ then satisfies
the null energy condition provided $\mu$ is chosen sufficiently large.
This is similar to the shock wave in the
Schwarzschild black hole case considered in~\cite{HPS2},
where a sufficiently large $\mu$ makes the null energy condition
hold everywhere except possibly near the singularity
where the linearised theory becomes unreliable.
Within our linearised treatment, the Bogoliubov coefficients
for the amended supertranslated wave in Rindler
would contain a new additive term coming from~$\mu$.
The explicit form of the correction would need to be determined
by an analysis similar to that in Section~\ref{BG}.

We anticipate that a similar analysis can be carried out for a 
shock wave that implants supertranslations on a Schwarzschild 
black hole~\cite{HPS2}, as shown in Figure~\ref{fig:schsup}, 
leading to non-trivial 
Bogoliubov
coefficients in the region $v > v_0$ outside the black hole.
There are now pairs of Hawking quanta created near the horizon, 
depicted as the pair $A \leftrightarrow B$ and the pair 
$C \leftrightarrow D$ in the figure, such that $A$ and $C$ 
are behind the Killing horizon while $B$ and $D$ are 
outgoing modes which an asymptotic observer at infinity 
will eventually detect as Hawking radiation. Each of the 
pairs $A \leftrightarrow B$ and $C \leftrightarrow D$ is 
maximally entangled. The infalling shock wave will then 
affect the entanglement between the interior quanta and 
the escaping quanta very much as in our Rindler analysis, 
so that the shock wave imprints its information on the 
Hawking quanta as a quantum memory. This may counteract 
the conventional argument that any 
characteristic information about infalling matter or 
radiation is lost in the Hawking evaporation~\cite{marolf,waldunruh}, 
and it may have a role in the proposal that supertranslations 
provide a solution to the 
black hole information paradox \cite{HPS1,HPS2,Strominger:2017aeh}
and in 
establishing a quantum version of black hole hair theorems.

\section*{Acknowledgments}

We thank Gerardo Adesso and Bartosz Regula for a helpful 
discussion on entanglement monotones and Masahiro Hotta for 
bringing the work in \cite{hotta1,hotta2} to our attention.
SK thanks the University of Nottingham 
and CPT, Aix Marseille Universit\'e, Universit\'e de Toulon, CNRS for hospitality and
the Department of Science and Technology, India, for partial
financial support. 
JL was supported in part by the 
Science and Technology Facilities Council 
(Theory Consolidated Grants ST/J000388/1 and ST/P000703/1).

\appendix

\section{Appendix: Bessel integral identity}\label{app:K-Bessel-integral}

The evaluation of the Bogoliubov coefficients in Section \ref{BG} uses the identity
\begin{align}
\int_0^\infty \frac{dx}{x} K_{i\Omega}(ax) K_{i\Omega'}(bx) 
&= 
\frac{\pi^2 {(b/a)}^{i\Omega}}{2\Omega\sinh(\pi\Omega)}
\left[1 + \frac{iq\Omega}{2} \, 
{}_2F_1 (1+i\Omega, 1; 2; q )
\right] 
\delta(\Omega-\Omega') 
\notag
\\[1ex]
&\hspace{3ex}
+ \frac{\pi^2 q {(b/a)}^{i\Omega'}}{8} 
{}_2F_1 \! \left(1+\frac{i(\Omega+\Omega')}{2}, 1 + \frac{i(\Omega'-\Omega)}{2}; 2; q \right)
\notag
\\
&\hspace{6ex}
\times \frac{1}{\sinh\bigl(\pi(\Omega+\Omega')/2\bigr)}
P\left(\frac{1}{\sinh\bigl(\pi(\Omega-\Omega')/2\bigr)}\right)
\ , 
\label{eq:Bessel-identity}
\end{align}
where $a>0$, $b>0$, $\Omega >0$, $\Omega' >0$, 
$q = 1 - b^2/a^2$, 
${}_2F_1$ is the Gaussian hypergeometric function \cite{dlmf}
and $P$ denotes the Cauchy principal value. 

To verify~\eqref{eq:Bessel-identity}, let $\epsilon>0$. 
We then have  
\begin{align}
\int_0^\infty \frac{dx}{x^{1-\epsilon}} K_{i\Omega}(ax) K_{i\Omega'}(bx) 
&= 
\frac{{(b/a)}^{i\Omega'} a^{-\epsilon}}{2^{3-\epsilon} \, \Gamma(\epsilon)} 
\left|\Gamma\left(\frac{\epsilon + i(\Omega+\Omega')}{2}\right)
\Gamma\left(\frac{\epsilon + i(\Omega-\Omega')}{2}\right)\right|^2
\notag
\\[1ex]
&\hspace{3ex}
\times 
{}_2F_1 \! \left(\frac{\epsilon + i(\Omega+\Omega')}{2}, \frac{\epsilon + i(\Omega'-\Omega)}{2}; 
\epsilon; q \right)
\notag 
\\
&= \frac{Q_\epsilon \pi^2{(b/a)}^{i\Omega'} (\Omega - \Omega')}{2 (\Omega+\Omega') 
\sinh\bigl(\pi(\Omega+\Omega')/2\bigr) 
\sinh\bigl(\pi(\Omega-\Omega')/2\bigr)} 
\notag 
\\[1ex]
&\hspace{3ex}
\times 
\frac{\epsilon}{(\Omega-\Omega')^2 + \epsilon^2} \, 
{}_2F_1 \! \left(\frac{\epsilon + i(\Omega+\Omega')}{2}, \frac{\epsilon + i(\Omega'-\Omega)}{2}; 
\epsilon; q \right)
\label{eq:Bessel-der1}
\end{align}
where $Q_\epsilon$ has the property that 
$Q_\epsilon\to1$ as $\epsilon\to0$. 
The first equality in \eqref{eq:Bessel-der1}
follows from formula 6.576.4 in \cite{Grad-Ryz}, 
and the second equality follows using standard 
properties of the Gamma-function~\cite{dlmf}. 

To evaluate the $\epsilon\to0$ limit in~\eqref{eq:Bessel-der1}, we expand 
${}_2F_1$ in its power series \cite{dlmf}
and use in each term the distributional identity 
\begin{align}
\lim_{\epsilon\to0_+}
\frac{1}{x\pm i\epsilon} 
= P \! \left(\frac{1}{x}\right) \mp i \pi \delta(x)
\ , 
\end{align}
with the outcome~\eqref{eq:Bessel-identity}.

\section{Appendix: Negativity}\label{app:negativity}

For all of our bipartite quantum systems, 
we quantify the entanglement by the negativity, 
defined by \cite{Peres-negativity,H3-negativity,vidal-werner:computable}
\begin{align}
{\mathcal N} = \tfrac12 \left( ||\rho^T|| - 1 \right) 
\ , 
\label{eq:neg-abstractdef}
\end{align} 
where $\rho$ is the density matrix, the superscript $T$ denotes the 
partial transpose, that is, the transpose in one of the subsystems, 
and $|| \cdot ||$ is the trace norm. 
An equivalent formula is 
\begin{align}
{\mathcal N} =\sum_i \tfrac12 \bigl( |\lambda_i| - \lambda_i \bigr)
\ , 
\label{eq:neg-sumdef}
\end{align} 
where $\lambda_i$ are the eigenvalues of $\rho^T$. 

${\mathcal N}$ is non-negative, and a strictly positive value 
of ${\mathcal N}$ implies that the system is not separable. 
${\mathcal N}$ does not in general coincide with the entanglement entropy, 
but it is an entanglement monotone, and although its 
operational meaning is subtle~\cite{plenio,audenart-plenio-eisert:cost}, 
it provides a convenient interpolation between other entanglement 
monotones with a more direct operational meaning~\cite{plenio-virmani:review}. 

The main advantage of ${\mathcal N}$ is that it is easy to compute 
in systems of arbitrary dimension. 
In this paper we consider applications to finite-dimensional Hilbert spaces; 
however, ${\mathcal N}$ generalises to infinite-dimensional Hilbert spaces, 
and it has a particularly convenient form for Fock state 
spaces in the continuous-variable formalism~\cite{Simon,Adesso}.


\end{document}